\documentclass{elsart}

 \usepackage{graphicx}

\usepackage{amssymb}

\usepackage{verbatim}

\newcounter{bla}
\newenvironment{refnummer}{%
\list{[\arabic{bla}]}%
{\usecounter{bla}%
 \setlength{\itemindent}{0pt}%
 \setlength{\topsep}{0pt}%
 \setlength{\itemsep}{0pt}%
 \setlength{\labelsep}{2pt}%
 \setlength{\listparindent}{0pt}%
 \settowidth{\labelwidth}{[9]}%
 \setlength{\leftmargin}{\labelwidth}%
 \addtolength{\leftmargin}{\labelsep}%
 \setlength{\rightmargin}{0pt}}}
 {\endlist}

\begin{document}
\begin{frontmatter}

\title{PLATYPUS: a code for reaction dynamics of weakly-bound nuclei at near-barrier energies within a classical dynamical model}

\author[a,b]{Alexis Diaz-Torres}


\address[a]{Department of Nuclear Physics, Research School of
Physical Sciences and Engineering, Australian National University,
Canberra, ACT 0200, Australia}
\address[b]{Department of Physics, University of Surrey, Guildford, GU2 7XH, UK}

\begin{abstract}
A self-contained Fortran-90 program based on a three-dimensional classical dynamical reaction model with stochastic breakup is presented, which is a useful tool for quantifying complete and incomplete fusion, and breakup in reactions induced by weakly-bound two-body projectiles near the Coulomb barrier. The code calculates (i) integrated complete and incomplete fusion cross sections and their angular momentum distribution, (ii) the excitation energy distribution of the \emph{primary} incomplete-fusion products, (iii) the asymptotic angular distribution of the incomplete-fusion products and the surviving breakup fragments, and (iv) breakup observables, such as angle, kinetic energy and relative energy distributions.     

\begin{flushleft}
PACS: 25.60.Pj, 25.60.Gc, 25.60-t, 24.10.-i

\end{flushleft}

\begin{keyword}
Complete fusion cross section, Incomplete fusion cross section, Breakup cross section, Spin distribution, Kinetic energy distribution, Relative energy distribution, Angle distribution, Classical trajectory, Monte Carlo sampling
\end{keyword}

\end{abstract}

\end{frontmatter}


{\bf PROGRAM SUMMARY/NEW VERSION PROGRAM SUMMARY}

\begin{small}
\noindent
{\em Manuscript Title:} PLATYPUS: a code for reaction dynamics of weakly-bound nuclei at near-barrier energies within a classical dynamical model
\\
{\em Authors:} Alexis Diaz-Torres                                             
\\
{\em Program Title:} PLATYPUS                                          
\\
{\em Journal Reference:}                                      \\
{\em Catalogue identifier:}                                   \\
{\em Licensing provisions:}                                   \\
{\em Programming language:} Fortran-90                        \\
{\em Computer:} 
Any Unix/Linux workstation or PC with a Fortran-90 compiler. \\
{\em Operating system:} Linux or Unix                         \\
{\em RAM:} 10 MB                                              \\
{\em Keywords:} Complete fusion cross section, Incomplete fusion cross section, Breakup cross section, Spin distribution, Kinetic energy distribution, 
Relative energy distribution, Angle distribution, Classical trajectory, Monte Carlo sampling \\
{\em PACS:} 25.60.Pj, 25.60.Gc, 25.60-t, 24.10.-i      \\
{\em Classification:}                                         \\
{\em External routines/libraries:} \\
  Several source routines from Numerical Recipies, and the Mersenne Twister 
  random number generator package are included to enable 
  independent compilation.                              
\\
{\em Nature of problem:} \\
  The program calculates a wide range of observables in reactions induced by 
  weakly-bound two-body nuclei near the Coulomb barrier. These include integrated 
  complete and incomplete fusion cross sections and their spin distribution, as well as 
  breakup observables (e.g. the angle, kinetic energy, and relative energy 
  distributions of the fragments).
   \\
{\em Solution method:} \\
  All the observables are calculated using a three-dimensional classical dynamical model 
  combined with the Monte Carlo sampling of probability-density distributions. 
  See Ref. [1,2] for further details.
   \\
{\em Restrictions:}\\
   The program is suited for a weakly-bound two-body projectile colliding with a stable 
   target. The initial orientation of the segment joining the two breakup fragments is 
   considered to be isotropic. \\
{\em Running time:}\\
   About 75 minutes for input provided, using a PC with 1.5 GHz processor.
   \\
{\em References:}
\begin{refnummer}
 \item A. Diaz-Torres et al., Phys. Rev. Lett. \textbf{98} (2007) 152701.
 \item A. Diaz-Torres, J. Phys. G: Nucl. Part. Phys. \textbf{37} (2010) 075109.

\end{refnummer}

\end{small}


\section{Introduction}

\begin{figure}
\begin{center}
\includegraphics[width=12.0cm,angle=0]{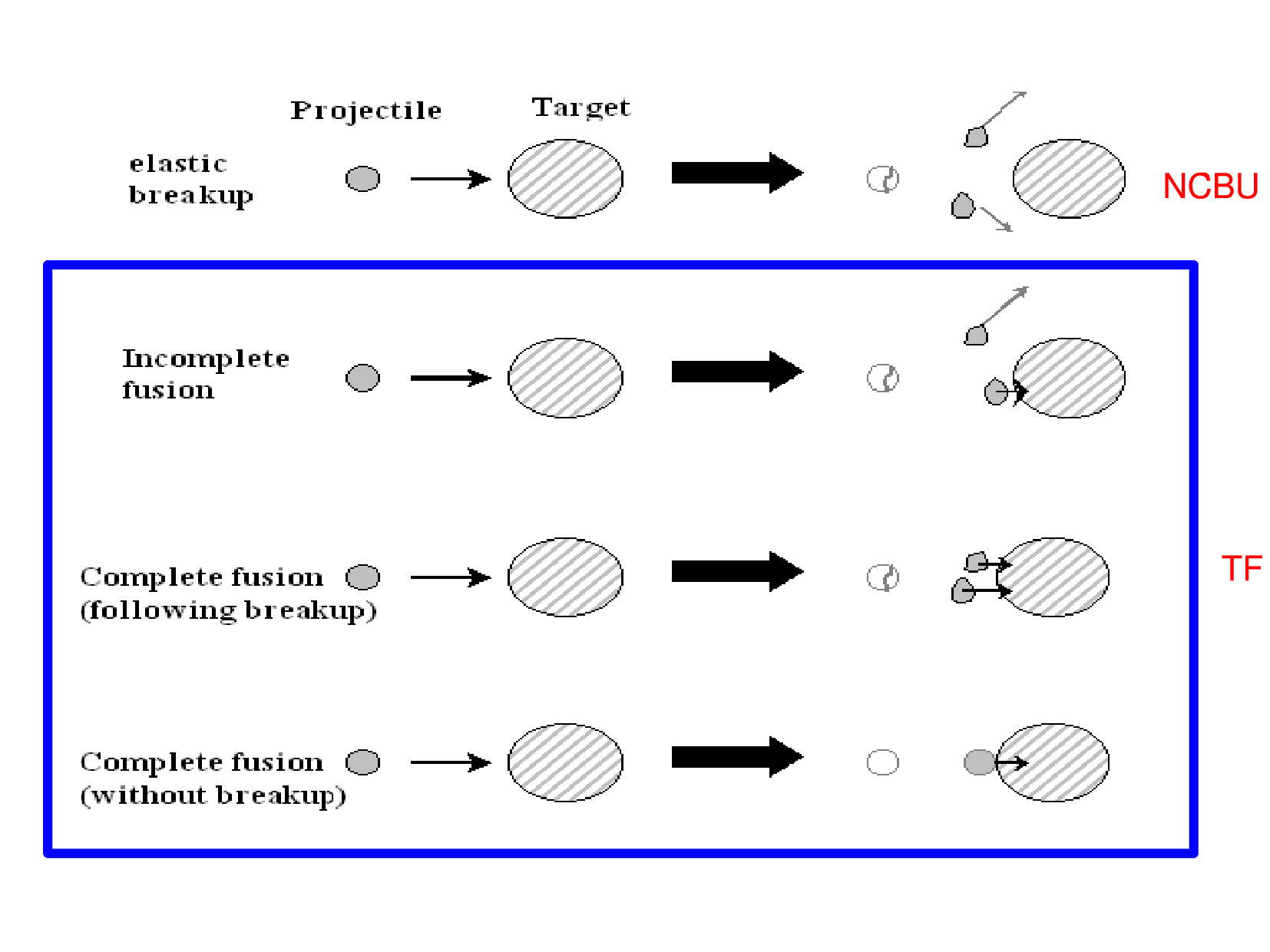}
\caption{Schematic of some relevant reaction processes of a two-body, weakly-bound projectile colliding with a stable target. The no-capture breakup ({\sc ncbu}) process and different components of the total fusion ({\sc tf}) process are highlighted. 
} 
\label{Figure1}
\end{center}
\end{figure}

Nuclear physics research has entered a new era with developments of rare-isotope beam facilities, where investigations are highly focused on understanding astrophysically important reaction rates involving exotic nuclei. These are often weakly-bound with a few-body, cluster structure that can easily be dissociated in their interaction with other nuclei. Understanding the breakup mechanism and its impact on nuclear reaction dynamics is essential. A major consequence of breakup is that a rich scenario of reaction pathways arises, such as events where (i) not all the resulting breakup fragments might be captured by the target, termed incomplete fusion ({\sc icf}), (ii) the entire projectile is captured by the target, called complete fusion ({\sc cf}), and (iii) none of the breakup fragments are captured, termed no-capture breakup ({\sc ncbu}), which is expected to be predominant at energies below the fusion barrier. Some of these are presented in 
Fig.\ \ref{Figure1}. Although it is not illustrated there, the transfer process is also very important \cite{Navin,Shrivastava,Chatterjee,Ramin,Huy}.

The experimental disentanglement of all these reaction processes is very complex. Also their modelling within a unified framework is an outstanding problem. The continuum-discretised coupled-channels (CDCC) model, for instance, can make reliable predictions of the {\sc ncbu} and {\sc tf} processes. However, this approach and other existing quantum models have limitations \cite{Thompson}, as they cannot calculate integrated incomplete and complete fusion cross sections unambiguously \cite{Alexis00,Alexis000}. Neither, after the formation of incomplete fusion products, can these follow the evolution of the surviving breakup fragment(s) since incomplete fusion results in depletion of the total few-body wave-function. A quantum model is very desirable, as it can deal with quantum tunnelling that is essential for understanding astrophysical reaction rates involving exotic nuclei. Nevertheless, some difficulties are overcome by the present classical dynamical reaction model.

This paper describes a computer program of the three-dimensional classical dynamical reaction model with stochastic breakup, recently published in Refs. \cite{Alexis0,Alexis1}. This approach exploits concepts and techniques (e.g., classical trajectory, Monte Carlo sampling) that overlap with some involved in dynamical models of multi-fragmentation and {\sc icf} in heavy-ion induced reactions at energies well-above the Coulomb barrier ($ \gtrsim 10$ MeV/nucleon), as in Refs. \cite{Moehring,Bondorf}. A crucial \emph{input} of the present model is a stochastically sampled breakup function suggested in Ref. \cite{Hinde}, which can be determined from sub-barrier breakup measurements \cite{Ramin,Hinde}. This function encodes the effects of the Coulomb and nuclear interactions that cause the projectile breakup. Hence, this approach is \emph{not} a breakup model, rather it is a quantitative dynamical model for relating the sub-barrier {\sc ncbu} to the above-barrier {\sc icf} and {\sc cf} of weakly-bound nuclei \cite{Hinde}. The model has successfully been applied to interpreting fusion and breakup measurements of weakly-bound nuclei \cite{Ramin,Huy,Santra}, and isomer ratio measurements \cite{Gasques}. Other applications in $\gamma$-ray spectroscopy are discussed in Ref. \cite{Alexis1}. 

Since the code is very user-friendly and useful for researchers involved in fusion and breakup measurements of weakly-bound nuclei at near-barrier energies, it will be beneficial make the program accessible to everyone. In Section 2 the program and the input file are explained. The code is illustrated in Section 3 with the reaction of a pseudo-$^{8}$Be projectile (assuming a weakly-bound state of two $\alpha$-particles \cite{Alexis0}) with a $^{208}$Pb target. Very recent measurements \cite{Ramin} have shown that prompt 
$^9$Be breakup occurs dominantly through an excited $^8$Be nucleus, validating the approximation of a $^9$Be projectile by $^8$Be. The good agreement of the classical model calculations for the {\sc ncbu} process with those of the CDCC quantum mechanical model shows the reliability of the classical dynamical approach (see Ref. \cite{Alexis0} for details). 

\section{Computer program and input file}
\subsection{Structure of the code}

The code has a main program and ten modules. The main program \textit{breakup3D} directs the input to be read, the problem to be solved, and details of the calculation to be written in output files. The modules are \textit{kinds, global data, potentials, mt19937, nrutil, initial conditions, fusion, angular momentum distribution, input values} and \textit{incomplete fusion products}.  

The main program \textit{breakup3D} calls the module \textit{input values} first, in which the subroutine \textit{input data} reads the input file described below. For a given partial wave between the projectile and the target (IMPACTMIN up to IMPACTMAX), the subroutines \textit{projectile trajectory} and \textit{trajectory arrays} of this module calculate the orbit of the bound projectile and store it for interpolations in the module \textit{initial conditions}. Thereafter, ISEEDMAX breakup events with sampled initial conditions are calculated for every partial wave. The initial conditions for the propagation in time of the three bodies are fixed by the subroutine \textit{initial values} of the module \textit{initial conditions}, whilst the classical trajectory of the breakup fragments and the target are calculated by the driver-subroutine \textit{ODEINT} of the module \textit{nrutil}. During the time propagation, the possible capture of the breakup fragments by the target is determined by the subroutine \textit{fusion events} in the module \textit{fusion}. Here, the relative energy and angular momentum between the nuclei, and the initial conditions for the time propagation in the two {\sc icf} channels are calculated as well. The spin distribution and cross sections for {\sc cf} and {\sc icf}, and for the breakup process are calculated by the subroutine \textit{spin distribution} in the module \textit{angular momentum distribution}. Here, other observables such as the angle, kinetic energy and relative energy distributions of the fragments from {\sc ncbu} events are also computed. Finally, the asymptotic angular distribution of the {\sc icf} products and the surviving breakup fragments is calculated by the subroutine \textit{propagating icf product} in the module \textit{incomplete fusion products}. This subroutine also provides the excitation energy distribution of the primary {\sc icf} products. Depending on the value of the output control variables (FILE1, FILE2, FILE3, FILE4, FILE5 and FILE6), details of the calculation can be written into output files. The output file TRACKING CALCULATIONS contains information on the evolution of the 
calculations. 

Module \textit{kinds}. This defines the kind type parameter for real values.

Module \textit{global data}. This defines global variables used in different modules.

Module \textit{potentials}. Here the nuclear and Coulomb interactions between the participants of the reaction (projectile fragments and the target) are defined.

Module \textit{mt19937}. It contains the Mersenne Twister random number generator, written by Makoto Matsumoto and Takuji Nishimura \cite{MTw}.

Module \textit{nrutil}. In this module several subroutines from Numerical Recipies are included, which are mainly related to integrating the classical equations of motion. \textit{ODEINT} is the integrator driver that calls the subroutine \textit{RKQC} which is a fouth-order Runge-Kutta integrator ensuring accuracy and adjusted stepsize. The forces are defined in the subroutine \textit{DERIVS}. The module also contains subroutines (\textit{printout1} and \textit{printout2}) for writing into output files the trajectory of the nuclei along with the value of the integrals of motion. The latter is crucial for checking the accuracy of the integration of the classical equations of motion.  

Module \textit{initial conditions}. This module defines the initial conditions for the propagation in time of the three bodies, following the breakup of the two-body projectile. This task is performed by the subroutine \textit{initial values}, which is guided by the conservation of energy, linear momentum and angular momentum in the overall center-of-mass frame of the projectile and target system. This is transformed to the laboratory reference frame (with Galilean kinematical relations) where the equations of motion are solved using a system of spherical coordinates. This module calls the module \textit{mt19937} in the sampling of (i) the initial excitation energy of the projectile, (ii) the initial relative angular momentum between the projectile fragments, (iii) the initial separation between the fragments, (iv) the breakup radius, (v) the initial orientation of the segment joining the two fragments of the projectile, and (vi) the initial direction of the radial velocity along that segment. If the output control variables are activated, the initial conditions will be written into output files. 

Module \textit{fusion}. In this module the subroutine \textit{fusion events} analyses the possible capture of any of the breakup fragments by the target. Here, the relative energy and angular momentum between the nuclei during the time propagation are calculated by the subroutines \textit{relative energy} and \textit{relative spin}. The subroutine \textit{icf product} calculates the initial positions and velocities for the time propagation in the two {\sc icf} channels.

Module \textit{angular momentum distribution}. Knowing the statistics of fusion and {\sc ncbu} events after a large number of sampling (1000 per partial wave in the example given below), the subroutine \textit{spin distribution} calculates observables related to {\sc cf} and {\sc icf} processes (spin distribution and cross sections). The angle, kinetic energy and relative energy distributions of the {\sc ncbu} events and the breakup cross section are also computed here. This subroutine extensively calls the subroutine \textit{LOCATE} of the module \textit{nrutil}. All observables are written into output files.  

Module \textit{input values}. Here the input file is read by the subroutine \textit{input data}. It also calculates the s-barrier features (radius and height) between (i) the projectile and the target, (ii) the projectile fragments and the target, and (iii) the two fragments of the projectile. The module also contains the subroutines \textit{projectile trajectory} and \textit{trajectory arrays} mentioned above.

Module \textit{incomplete fusion products}. This module includes the subroutines \textit{propagating icf product} and \textit{icf trajectory}. The former carries out the time propagation of the {\sc icf} products and the remaining breakup fragments, whilst the latter analyses this propagation in order to determine their asymptotic angles. The angular distributions, and the excitation energy distribution of the primary {\sc icf} products are constructed like those in the module \textit{angular momentum distribution}. All observables are also written into output files. 

\subsection{Input file}

Fig. 2 shows the input file of the program. At the bottom of the figure, the namelist of the input variables appearing in the code are shown. The lines related to the potentials (lines 12--21) are self-explanatory, whilst some variables in previous lines have already been mentioned in the description of the program. We will here describe lines 1--11 only. 

\begin{verbatim}
0,0,0,0,0,0        
0,45,1000     
0                 
50.              
6.,4.        
1.05,0.        
1,0.922           
1.80,1.        
0.922,9.73      
50.
0             
#######POTENTIALS################
208.,82.        
8.,4.        
4.,2.        
4.,2.        
-120.9030,1.39,0.7552,1.2  
-62.,1.39,0.62,1.2  
-62.,1.39,0.62,1.2  
-16.6964,1.2,0.62,1.2
-33.9757,1.4788,0.63,1.2,-8.95  
-33.9757,1.4788,0.63,1.2,-8.95  
#################################
Line 1: FILE1, FILE2, FILE3, FILE4, FILE5, FILE6
Line 2: IMPACTMIN, IMPACTMAX, ISEEDMAX
Line 3: PARTICULAR_IMPACT 
Line 4: E0
Line 5: EXCMAX, L12MAX
Line 6: EXCMIN, L12MIN
Line 7: TYPE_EXC, alfaexc (only important when TYPE_EXC=1)
Line 8: d012,sig012
Line 9: alpha, beta 
Line 10: RBU_max
Line 11: ICF_PROPAG
Line 12: Mass and charge of the target (AT,ZT)
Line 13: Mass and charge of the projectile (AP,ZP)
Line 14: Mass and charge of the fragment1 (AP1,ZP1)
Line 15: Mass and charge of the fragment2 (AP2,ZP2)  
Line 16: Target-Projectile potential (V0TP,rr0TP,a0TP) 
         and Coulomb radius (rrc0TP).
         *********************************************** 
         * In the potential, the radius parameters are * 
         * multiplied by AT**(1/3)                     *
         *********************************************** 
Line 17: Target-Fragment1 potential (V01,rr01,a01) and 
         and Coulomb radius (rrc01)
         *********************************************** 
         * In the potential, the radius parameters are * 
         * multiplied by AT**(1/3)                     *
         ***********************************************   
Line 18: Target-Fragment2 potential (V02,rr02,a02) and 
         and Coulomb radius (rrc02) 
         *********************************************** 
         * In the potential, the radius parameters are * 
         * multiplied by AT**(1/3)                     *
         ***********************************************
Line 19: Fragment1-Fragment2 potential (V012,rr012,a012) 
         and Coulomb radius (rrc012)
         *********************************************** 
         * In the potential, the radius parameters are * 
         * multiplied by AP1**(1/3)                    *
         ***********************************************
Line 20: Potential for ICF channel 1: (AT+AP1) and AP2
         (V01ICF,r01ICF,a01ICF), Coulomb radius (rc1ICF)
         and Qvalue (Qvalue1) for the ICF process
         *********************************************** 
         * In the potential, the radius parameters are * 
         * multiplied by (AT+AP1)**(1/3)               *
         ***********************************************
Line 21: Potential for ICF channel 2: (AT+AP2) and AP1
         (V02ICF,r02ICF,a02ICF), Coulomb radius (rc2ICF)
         and Qvalue (Qvalue2) for the ICF process
         *********************************************** 
         * In the potential, the radius parameters are * 
         * multiplied by (AT+AP2)**(1/3)               *
         ***********************************************
\end{verbatim}
\begin{center}
Fig. 2. Input file for PLATYPUS code (PLATYPUS.inp).
\end{center}

The integer variables of the first line allow the user to write details of the dynamical calculations into output files, when they are equal to one. FILE1 and FILE2 open output files to write the trajectory of the breakup fragments and the initial projectile-target, respectively. The file opened by FILE3 contains details of breakup events, and FILE4 is for plotting the trajectory of the breakup fragments. FILE5 opens a file to write the asymptotic angles and energy of the breakup fragments, and FILE6 provides a file for the asymptotic angles of the {\sc icf} products and the surviving breakup fragments. Line 2 defines the window of orbital angular momentum of the incident projectile (IMPACTMIN, IMPACTMAX) in units of $\hbar$, and the number of breakup events per partial wave (ISEEDMAX). PARTICULAR IMPACT in line 3 is to select a partial wave, whose associated dynamical calculations are written into the file opened by FILE3. Line 4 refers to the incident energy of the projectile in MeV (E0). Lines 5--6 define the range of initial excitation energy (EXCMIN, EXCMAX) in MeV and relative angular momentum (L12MIN, L12MAX) in units of $\hbar$ for the breakup fragments. In line 7, TYPE EXC controls the sampling function for excitation where for TYPE EXC=0 the weighting is uniform, and for TYPE EXC=1 it is exponentially decreasing with a coefficient alfaexc. Line 8 defines the centroid (d012) and width (sig012) in fm of the Gaussian function that describes the radial probability distribution of projectile ground-state wave function. It is used to sample the initial separation between the breakup fragments. The breakup function [see expression (1) in Refs. \cite{Alexis0,Alexis1}] is given in line 9 by the parameters alpha and beta, being beta=$\ln (A)$. Line 10 defines a maximal projectile-target separation for sampling the breakup radius, whilst line 11 allows the user to carry out the time propagation in the {\sc icf} channels, when ICF PROPAG=1.

\section{Test run}

Using the input file of Fig. 2, Figs. 3--5 show output files for angular momentum distribution in {\sc icf}, {\sc cf} and {\sc ncbu} processes, respectively. The angular momenta are in the first column, whilst in the second and third columns are partial probabilities and cross sections. The total cross section appears at the end of the file. 

\begin{verbatim}
REGARDING ANGULAR MOMENTUM BROUGHT BY FRAGMENTS INTO TARGET

L(hbar)  P_L      SIGMA_L (mb)

  0.00  .475E-01  0.8469E+00
  1.00  .303E+00  0.6309E+01
  2.00  .573E+00  0.1412E+02
  3.00  .770E+00  0.2075E+02
  4.00  .915E+00  0.2678E+02
  5.00  .977E+00  0.2914E+02
  6.00  .965E+00  0.3215E+02
  7.00  .933E+00  0.3361E+02
  8.00  .822E+00  0.3315E+02
  9.00  .739E+00  0.3322E+02
 10.00  .503E+00  0.2349E+02
 11.00  .354E+00  0.1667E+02
 12.00  .193E+00  0.9085E+01
 13.00  .941E-01  0.4320E+01
 14.00  .549E-01  0.2465E+01
 15.00  .239E-01  0.1121E+01
 16.00  .191E-01  0.8695E+00
 17.00  .461E-02  0.2010E+00
 18.00  .659E-03  0.3332E-01

TOTAL SIGMA_ICF = 0.2883E+03 mb

REGARDING PROJECTILE ANGULAR MOMENTUM

  0  0.3230E+00  0.5266E+00
  1  0.3323E+00  0.1625E+01
  2  0.3474E+00  0.2832E+01
  3  0.3415E+00  0.3897E+01
  4  0.3310E+00  0.4856E+01
  5  0.3659E+00  0.6561E+01
  6  0.3764E+00  0.7978E+01
  7  0.3929E+00  0.9608E+01
  8  0.4127E+00  0.1144E+02
  9  0.4279E+00  0.1325E+02
 10  0.4536E+00  0.1553E+02
 11  0.4819E+00  0.1807E+02
 12  0.5043E+00  0.2055E+02
 13  0.5281E+00  0.2324E+02
 14  0.5281E+00  0.2497E+02
 15  0.5162E+00  0.2609E+02
 16  0.4424E+00  0.2380E+02
 17  0.3995E+00  0.2280E+02
 18  0.3659E+00  0.2207E+02
 19  0.1681E+00  0.1069E+02
 20  0.1112E+00  0.7432E+01
 21  0.6120E-01  0.4290E+01
 22  0.3758E-01  0.2757E+01
 23  0.2215E-01  0.1697E+01
 24  0.1191E-01  0.9517E+00
 25  0.5000E-02  0.4157E+00
 26  0.2346E-02  0.2027E+00
 27  0.6873E-03  0.6162E-01
 28  0.8565E-03  0.7959E-01
 29  0.2373E-03  0.2282E-01
 30  0.1314E-03  0.1307E-01
 31  0.1638E-03  0.1682E-01
 33  0.7519E-04  0.8213E-02

TOTAL SIGMA_ICF = 0.2883E+03 mb
\end{verbatim}
\begin{center}
Fig. 3. {\sc icf} angular momentum distribution (ICF SPIN DISTRIBUTION).
\end{center}

In Fig. 3, the {\sc icf} spin distribution refers to either the angular momentum brought by the captured fragment into the target or the orbital angular momentum of the incident projectile. The former has been employed very recently to understand isomer ratio measurements \cite{Gasques}. Of course, both distributions are the same for {\sc cf}. 

\begin{verbatim}
REGARDING ANGULAR MOMENTUM BROUGHT BY FRAGMENTS INTO TARGET

L(hbar)  P_L      SIGMA_L (mb)

  0.00  .334E+00  0.5438E+00
  1.00  .326E+00  0.1596E+01
  2.00  .311E+00  0.2531E+01
  3.00  .316E+00  0.3611E+01
  4.00  .326E+00  0.4778E+01
  5.00  .291E+00  0.5214E+01
  6.00  .280E+00  0.5924E+01
  7.00  .264E+00  0.6448E+01
  8.00  .242E+00  0.6705E+01
  9.00  .222E+00  0.6882E+01
 10.00  .194E+00  0.6636E+01
 11.00  .164E+00  0.6155E+01
 12.00  .133E+00  0.5401E+01
 13.00  .956E-01  0.4208E+01
 14.00  .659E-01  0.3117E+01
 15.00  .442E-01  0.2232E+01
 16.00  .356E-01  0.1915E+01
 17.00  .244E-01  0.1392E+01
 18.00  .138E-01  0.8351E+00
 19.00  .396E-02  0.2515E+00

TOTAL SIGMA2_CF = 0.7637E+02 mb

SIGMA_CF(INERT PROJ) = 0.6521E+03 mb
SIGMA2_CF =   0.7637E+02 mb
SIGMA_CF(NO BU) =   0.2222E+03 mb
SIGMA_CF(TOT) =   0.2986E+03 mb

****** Terminology ******

SIGMA_CF(INERT PROJ): when the projectile cannot break up.
SIGMA2_CF: complete fusion following breakup.
SIGMA_CF(NO BU): from events that survive the breakup.
SIGMA_CF(TOT): SIGMA_CF(NO BU) + SIGMA2_CF
\end{verbatim}
\begin{center}
Fig. 4. {\sc cf} angular momentum distribution (CF SPIN DISTRIBUTION).
\end{center}

Since the component of the {\sc cf} spin distribution associated with the fusion of the bound projectile is easily calculated in terms of the projectile critical angular-momentum for fusion (see first term of expression (4) in Refs. \cite{Alexis0,Alexis1}), in Fig. 4 we highlight only the nontrivial component related to the capture of all the projectile fragments after breakup. The code also calculates {\sc cf} cross sections related to the inert projectile and to the mentioned two components of this process when the projectile can be dissociated, as shown at the end of Fig. 4. As expected, the {\sc ncbu} process in Fig. 5 reveals a broader angular momentum distribution than the {\sc cf} and {\sc icf} processes. 

\begin{verbatim}
REGARDING PROJECTILE ANGULAR MOMENTUM

L(hbar)  PBU_L      SIGMA_L (mb)

  0  0.2637E-02  0.4299E-02
  1  0.6593E-03  0.3224E-02
  2  0.1319E-02  0.1075E-01
  3  0.1319E-02  0.1505E-01
  4  0.2637E-02  0.3869E-01
  5  0.2637E-02  0.4729E-01
  6  0.3296E-02  0.6986E-01
  7  0.2637E-02  0.6448E-01
  8  0.4615E-02  0.1279E+00
  9  0.9230E-02  0.2859E+00
 10  0.1187E-01  0.4063E+00
 11  0.1319E-01  0.4944E+00
 12  0.2242E-01  0.9135E+00
 13  0.3560E-01  0.1567E+01
 14  0.6527E-01  0.3086E+01
 15  0.9889E-01  0.4998E+01
 16  0.1813E+00  0.9753E+01
 17  0.2354E+00  0.1343E+02
 18  0.2795E+00  0.1686E+02
 19  0.4661E+00  0.2963E+02
 20  0.3738E+00  0.2498E+02
 21  0.3064E+00  0.2148E+02
 22  0.2564E+00  0.1881E+02
 23  0.2131E+00  0.1633E+02
 24  0.1819E+00  0.1453E+02
 25  0.1548E+00  0.1287E+02
 26  0.1288E+00  0.1113E+02
 27  0.1068E+00  0.9572E+01
 28  0.8926E-01  0.8295E+01
 29  0.7434E-01  0.7151E+01
 30  0.6165E-01  0.6131E+01
 31  0.5137E-01  0.5276E+01
 32  0.4256E-01  0.4510E+01
 33  0.3448E-01  0.3766E+01
 34  0.2913E-01  0.3277E+01
 35  0.2410E-01  0.2789E+01
 36  0.1998E-01  0.2377E+01
 37  0.1674E-01  0.2046E+01
 38  0.1367E-01  0.1716E+01
 39  0.1130E-01  0.1456E+01
 40  0.9314E-02  0.1230E+01
 41  0.7709E-02  0.1043E+01
 42  0.6334E-02  0.8777E+00
 43  0.5261E-02  0.7462E+00
 44  0.4334E-02  0.6289E+00
 45  0.3512E-02  0.5210E+00

SIGMA_BU = 0.2653E+03 mb
\end{verbatim}
\begin{center}
Fig. 5. {\sc ncbu} angular momentum distribution (BU SPIN DISTRIBUTION).
\end{center}


%


\newpage
\textbf{Acknowledgement}
I thank R. Rafiei, D.H. Luong, M. Brown, J.A. Tostevin, D.J. Hinde, M. Dasgupta and L.R. Gasques for discussions and constructive suggestions. The work was supported by an ARC Discovery Grant and the UK Science and Technology Facilities Council (STFC) Grant No. ST/F012012/1.


\begin{thebibliography}{00}




\bibitem{Dasgupta0} M.~Dasgupta et al., Phys. Rev. Lett. \textbf{82} (1999) 1395; 
Phys. Rev. C \textbf{70} (2004) 024606.
\bibitem{Beck0} C.~Beck et al., Phys. Rev. C \textbf{67} (2003) 054602; Nucl. Phys. A \textbf{834} (2010) 440c.
\bibitem{Navin} A.~Navin et al., Phys. Rev. C \textbf{70} (2004) 044601.
\bibitem{Shrivastava} A.~Shrivastava et al., Phys. Lett. B \textbf{633} (2006) 463.
\bibitem{Chatterjee} A.~Chatterjee et al., Phys. Rev. Lett. \textbf{101} (2008) 032701.
\bibitem{Ramin} R.~Rafiei et al., Phys. Rev. C \textbf{81} (2010) 024601.
\bibitem{Huy} D.~H.~Luong et al., Phys. Lett. B (2010), doi:10.1016/j.physletb.2010.11.007.
\bibitem{Thompson} I.~J.~Thompson, A.~Diaz-Torres, Prog. Theor. Phys. Suppl. \textbf{154} (2004) 69.
\bibitem{Alexis00} A.~Diaz-Torres, I.~J.~Thompson, Phys. Rev. C \textbf{65} (2002) 024606.
\bibitem{Alexis000} A.~Diaz-Torres, I.~J.~Thompson, C.~Beck, Phys. Rev. C \textbf{68} (2003) 044607. 
\bibitem{Alexis0} A.~Diaz-Torres et al., Phys. Rev. Lett. \textbf{98} (2007) 152701.
\bibitem{Alexis1} A.~Diaz-Torres, J. Phys. G: Nucl. Part. Phys. \textbf{37} (2010) 075109.
\bibitem{Moehring} K.~M\"ohring et al., Phys. Lett. B \textbf{203} (1988) 210.
\bibitem{Bondorf} J.~P.~Bondorf et al., Phys. Rev. C \textbf{46} (1992) 374.
\bibitem{Hinde} D.~J.~Hinde et al., Phys. Rev. Lett. \textbf{89} (2002) 272701.
\bibitem{Santra} S.~Santra et al. Phys. Lett. B \textbf{677} (2009) 139.
\bibitem{Gasques} L.~R.~Gasques et al., Phys. Rev. C \textbf{74} (2006) 064615. 
\bibitem{MTw} http://www.math.keio.ac.jp/matumoto/emt.html 

\end{thebibliography}
\end{document}